\documentclass[]{elsarticle}
\usepackage[margin=3cm]{geometry}
\usepackage{graphicx}
\usepackage{amsmath}
\usepackage{amssymb}

\begin{document}

\title{\Large  {{\bf{Spontaneous $R$-Parity Breaking, Stop LSP Decays \\
and the Neutrino Mass Hierarchy}}}}
\author{{Zachary Marshall${}^{1}$, Burt A.~Ovrut${}^{2}$, Austin Purves${}^{2}$ and Sogee Spinner${}^{2}$} \\[5mm]
    {\it  ${}^{1}$ Physics Division, Lawrence Berkeley National Laboratory} \\
   {\it Berkeley, CA 94704}\\[4mm]
    {\it  ${}^{2}$ Department of Physics, University of Pennsylvania} \\
   {\it Philadelphia, PA 19104--6396}\\[4mm]
}
\date{\today}
\begin{abstract}
\let\thefootnote\relax\footnotetext{\mbox{
ovrut@elcapitan.hep.upenn.edu,  ~apurves@sas.upenn.edu, 
~sogee@sas.upenn.edu, ~ zlmarshall@lbl.gov}}The MSSM with right-handed neutrino supermultiplets, gauged $B-L$ symmetry and a non-vanishing sneutrino expectation value is the minimal theory that spontaneously breaks $R$-parity and is consistent with the bounds on proton stability and lepton number violation. This minimal $B-L$ MSSM can have a colored/charged LSP, of which a stop LSP is the most amenable to observation at the LHC. We study the $R$-parity violating decays of a stop LSP into a bottom quark and charged leptons--the dominant modes for a generic ``admixture'' stop. A numerical analysis of the relative branching ratios of these decay channels is given using a wide scan over the parameter space. The fact that $R$-parity is violated in this theory by a vacuum expectation value of a sneutrino links these branching ratios directly to the neutrino mass hierarchy. It is shown how a discovery of bottom-charged lepton events at the LHC can potentially determine whether the neutrino masses are in a normal or inverted hierarchy, as well as determining the $\theta_{23}$ neutrino mixing angle. Finally, present LHC bounds on these leptoquark signatures are used to put lower bounds on the stop mass.
\end{abstract}
\maketitle
\newpage
\interfootnotelinepenalty=10000

\section*{Introduction}

The extension of the standard $SU(3)_{C} \times SU(2)_{L} \times U(1)_{Y}$ model of particle physics, with or without right-handed neutrinos, to $N=1$ supersymmetry (SUSY) is immediately confronted by a fundamental problem. Without any further constraints, the superpotential must contain cubic superfield interactions that violate both baryon number ($B$) and lepton number ($L$)--thus leading, at tree level, to potentially rapid proton decay and unobserved lepton number violating processes. The conventional ``natural'' solution to this problem is to demand that the Lagrangian be invariant under a discrete $R$-parity, $R=(-1)^{3(B-L)+2s}$ where $s$ is the spin of the component particle. This symmetry indeed eliminates the dangerous $B$ and $L$ violating interactions, and is consistent with the observed constraints on these quantities. The $R$-parity invariant supersymmetric extension of the standard $SU(3)_{C} \times SU(2)_{L} \times U(1)_{Y}$ model of particle physics, with or without right-handed neutrinos, is referred to as the minimal supersymmetric standard model (MSSM), and is the usual paradigm for a low energy supersymmetric particle physics model.

Be this as it may, from the low energy point of view the imposition of discrete $R$-parity is completely {\it ad hoc}. There have been many attempts to justify it by 1) embedding the MSSM into a supersymmetric grand unified theory (GUT), \textit{e.g.} \cite{Aulakh:2000sn}, or 2) as arising from a residual topological, finite or anomalous Abelian symmetry of a superstring vacuum \cite{Braun:2006me, Anderson:2010tc}. Without prejudice as to the efficacy or physical reality of these attempts, there is another way to arrive at the same results which is both straightforward, natural and does not require the introduction of any superfields beyond those of the MSSM with right-handed neutrino supermultiplets. This is as follows. 

It has been known for a long time that the right-handed neutrino version of the SM--and its MSSM extension--remains anomaly free if one enlarges the gauge group to $SU(3)_{C} \times SU(2)_{L} \times U(1)_{Y} \times U(1)_{B-L}$. Furthermore, note that $R$-parity is a discrete ${\mathbb{Z}}_{2}$ subgroup of $U(1)_{B-L}$. It follows that one can ``naturally'' incorporate $R$-parity conservation into the MSSM with right-handed neutrinos simply by extending the gauge group to $SU(3)_{C} \times SU(2)_{L} \times U(1)_{Y} \times U(1)_{B-L}$. 
However, since it is unobserved at the electroweak scale, this gauged $U(1)_{B-L}$ symmetry must be broken at, say, a TeV scale or above. There have been attempts to do this, while leaving $R$-parity unbroken. This can only be accomplished, however, by introducing new chiral multiplets with even $B-L$ charge \cite{Font:1989ai}. That is, one must go beyond the MSSM particle content and introduce new fields into the spectrum. However, one need not preserve $R$-parity if the scale of its breaking is sufficiently low--for example, at a TeV. This can be accomplished if one, or more, of the right-handed sneutrino scalars--each carrying an odd $B-L$ charge--develop a vacuum expectation value (VEV). This does not require the introduction of any additional multiplets and is consistent with proton stability--since a sneutrino VEV breaks lepton number only--and the bounds on lepton violation. 

We will refer to this theory as the minimal $B-L$ MSSM. It was introduced from the ``bottom up'' point of view in \cite{FileviezPerez:2008sx, Barger:2008wn, Everett:2009vy}\footnote{Such a minimal model was outlined as a possible low energy manifestation of $E_6$ GUT models in~\cite{Mohapatra:1986aw}.}. It was also found from a ``top down'' perspective to be the low energy theory associated with a class of vacua of $E_{8} \times E_{8}$ heterotic $M$-theory \cite{Lukas:1998yy, Braun:2005ux, Braun:2005nv, Braun:2006ae, Ambroso:2009jd}. Various aspects of this minimal theory were subsequently discussed, such as the radiative breakdown of the $U(1)_{B-L}$ gauge symmetry and its hierarchy with electroweak breaking \cite{Ovrut:2012wg, Ambroso:2009sc, Ambroso:2010pe}, the neutrino sector \cite{Mohapatra:1986aw, Ghosh:2010hy, Barger:2010iv}, possible LHC signals~\cite{FileviezPerez:2012mj, Perez:2013kla} and some cosmological effects~\cite{Perez:2013kla}. We take the point of view that this $B-L$ MSSM is the minimal possible extension of the MSSM that is consistent with proton stability and observed lepton violation bounds. Hence, it is potentially a realistic candidate for a low energy $N=1$ supersymmetric particle physics model. With this in mind, we wish to study the the dominant signatures of this model at the Large Hadron Collider (LHC) that can distinguish it from the MSSM. The initial results of this study are presented in this paper.

We find that there are three distinct phenomena that can occur in the minimal $B-L$ MSSM that are potentially observable at the LHC and sharply distinguish this model from the MSSM. These are the following.

\begin{itemize}

\item Since $R$-parity is violated in the minimal $B-L$ MSSM, it is now possible that the lightest supersymmetric particle (LSP)\footnote{Throughout this paper, we use the term LSP to refer to the lightest supersymmetric particle {\it relevant for collider physics}.} can carry color and/or electric charge without coming into conflict with astrophysical data. This is because the LSP can now decay sufficiently quickly via $R$-parity violating operators. Furthermore, the specific nature of this theory--which exactly specifies the $R$-parity violating vertices and their relative strengths--determines all LSP decay products and their branching ratios.

\item The ``Higgs'' field that spontaneously breaks $U(1)_{B-L}$ in this minimal model is at least one of the right-handed sneutrinos. It follows that the neutrino sector in this theory is intimately related to the $R$-parity violating operators and, hence, to the allowed decay products of the LSP and their branching ratios. Put the other way, observation at the LHC of the relative branching ratios of the LSP decays can directly inform specific issues in the neutrino mass matrix--specifically, whether there is a ``normal'' or an ``inverted'' neutrino mass hierarchy and can potentially remove the ambiguity in the measurement of the $\theta_{23}$ mixing angle, which can be one of two measured central values.

\item As mentioned above, the minimal $B-L$ theory exactly specifies the allowed $R$-parity violating decays of the LSP. For a chosen LSP, these decay signatures, which are disallowed within the $R$-parity invariant MSSM, can be rather unique. Data on such decays at the LHC can then be used to put a lower bound on the LSP mass. 

\end{itemize}

We hasten to point out that confirmation at the LHC of $R$-parity violating LSP decays consistent with the minimal $B-L$ MSSM is not sufficient to establish its reality. Full confirmation of this theory would require at least two other specific discoveries: 1) a massive vector boson in the TeV range corresponding to $B-L$ and 2) the existence of some other explicit superpartner. Be that as it may, a careful study of the three issues discussed in the bullet points--and their implications for the LHC--would be a major step in either confirming, putting bounds on, or disproving the minimal $B-L$ MSSM. We now present the results of such a study.
The technical details will be presented in a forthcoming publication \cite{Purves}.

\section*{$R$-Parity Violation and Stop LSP Decays}

First a technical point. It will be assumed in this paper that all  gauge couplings of the minimal $B-L$ MSSM unify at a high scale. Under this assumption, we find it easier to work with the rotated Abelian gauge groups $U(1)_{3R} \times U(1)_{B-L}$  rather than $U(1)_{Y} \times U(1)_{B-L}$, since the former, unlike the original gauge group, has no kinetic mixing at any scale. This greatly simplifies the calculations, while changing none of the physics conclusions.

It was shown in \cite{Ghosh:2010hy, Barger:2010iv} that within the minimal $B-L$ MSSM all non-vanishing right-handed sneutrino VEV's can, without loss of generality, be rotated into the third family, and that this VEV is given by
\begin{equation}
v_R^2=\frac{-8m^2_{\tilde \nu_{3}^c}  + g_R^2\left(v_u^2 - v_d^2 \right)}{g_R^2+g_{BL}^2} 
\label{B1}
\end{equation}
where $m_{{\tilde \nu_{3}}^c}$ and $v_{u}$, $v_{d}$ are the third family sneutrino soft SUSY breaking mass parameter and the up-, down-Higgs VEV's respectively. The parameters $g_{R}$ and $g_{BL}$ are the gauge couplings for $U(1)_{3R}$ and $U(1)_{B-L}$. Furthermore, $v_{R}$ induces a smaller VEV for each of the left-handed sneutrinos given by
\begin{equation}
{v_L}_i=\frac{\frac{v_R}{\sqrt 2}(\mu \, Y_{\nu_{i3}}^*v_d-a_{\nu_{i3}}^*v_u)}{m_{\tilde L_{i}}^2-\frac{g_2^2}{8}(v_u^2-v_d^2)-\frac{g_{BL}^2}{8}v_R^2}
\label{B2}
\end{equation}
for $i=1,2,3$. Here $Y_{\nu_{i3}}$ and $m_{\tilde L_{i}}$ are the neutrino $(i3)$-Yukawa couplings and the left-handed sneutrino soft SUSY breaking mass parameters respectively, $\mu$ is the mu-parameter, $a_{\nu_{i3}}$ are the $(i3)$-components of the sneutrino tri-linear soft SUSY breaking terms, and $g_{2}$ is the gauge coupling parameter for $SU(2)_{L}$. These expectation values spontaneously break the gauged $U(1)_{3R} \times U(1)_{B-L}$ symmetry down to $U(1)_{Y}$. When expanded around these VEV's, explicit $R$-parity violating terms appear in the Lagrangian. It is these terms that lead to decays of the LSP. These terms are similar to explicit bilinear $R$-parity violation in the MSSM, although there are important differences stemming from the neutrino sector. For example, bilinear $R$-parity violation has only one massive neutrino at tree level, whereas our model has two. For earlier works on non-LSP stop decays see~\cite{Diaz:1999ge, Restrepo:2001me,Datta:2006ak}. In addition, for a study on the relationship between neutrino masses and collider phenomenology in the MSSM with explicit trilinear $R$-parity violation, see~\cite{Barger:2001xe}.

Generically, within the minimal $B-L$ MSSM any superpartner can potentially be the LSP. Be that as it may, colored particles are more readily produced at the LHC and, hence, one can put more aggressive bounds on their decays. Furthermore, if one assumes unification of the gauge coupling parameters, then it was shown in \cite{Ovrut:2012wg} that the gluino cannot be the LSP. Therefore, one is driven to consider squark LSP's only. However, it is well-known from renormalization group analyses of the mass parameters \cite{Martin:1997ns} that the third family of squarks is generically the lightest. Hence, one should consider both the stop and the sbottom as potential LSP candidates. In this paper, we will, for simplicity, limit the discussion to a stop LSP, deferring the analysis of a sbottom LSP to a forthcoming paper.

The left stop-right stop mass matrix is a function of a number of parameters in the $B-L$ MSSM Lagrangian. This can be diagonalized into a light stop, denoted ${\tilde{t}}_{1}$, which we take to be the LSP and a heavier stop, ${\tilde{t}}_{2}$, which can henceforth be ignored. This LSP can be shown to always decay via $R$-parity violating interactions into a lepton and a quark--that is, ${\tilde{t}}_{1}$ behaves as a ``leptoquark''. Furthermore, if one only considers generic values of the left and right stop mixing angle, denoted by $\theta_{t}$--that is, ${\tilde{t}}_{1}$ is a generic admixture of the left and right stops and {\it not} purely a right stop-- then the decay into a bottom quark and a charged lepton dominates over the decay into a top quark and a neutrino. This latter decay will be neglected here, but discussed in detail in~\cite{Purves}.
\begin{equation}
{\tilde{t}}_{1} \longrightarrow b~ \ell^+_{i}~,~~i=1,2,3
\label{B3}
\end{equation}
where $b$ is the bottom quark and $\ell^+_{i}, i=1,2,3$, are the positron, anti-muon and anti-tau respectively.

The partial widths of a stop LSP into bottom--charged leptons can be calculated, and are found to be
\begin{equation}
\Gamma(\tilde t_1 \to b \, \ell^+_i)=\frac{1}{16\pi}(|G^L_{{\tilde t}_{1} b\ell_{i}}|^2+|G^{R}_{{\tilde t}_{1}b\ell_{i}}|^2)m_{\tilde t_1}
\label{B4}
\end{equation}
where, $G^L_{\tilde t_{1} b\ell_{i}}$ and $G^R_{\tilde t_{1} b\ell_{i}}$ are complicated functions of a large number of parameters in the $B-L$ MSSM Lagrangian and $m_{{\tilde t}_{1}}$ is the LSP mass. To illustrate this parameter dependence, we note that they can be approximated by
\begin{eqnarray}
	G^L_{\tilde t_1  b \ell_i} & =&-Y_b c_{\theta_t}
		\frac{1}{\mu} \epsilon_i \label{B5}
	\\
	G^R_{\tilde t_1  b \ell_i} & =& 
	-g_2^2 c_{\theta_t} \frac{\tan \beta m_{\ell_i}}{\sqrt 2 M_2 \mu} {v_L}_i^*
	- Y_t s_{\theta_t} \frac{m_{\ell_i}}{\sqrt 2 v_d \mu} {v_L}_i^* \label{B6} 
\end{eqnarray}
where $\epsilon_i = \frac{1}{\sqrt 2} {Y_\nu}_{i3} v_R $, $Y_b$ and $Y_{t}$ are the bottom and top quark Yukawa couplings respectively, $M_{2}$ is the $SU(2)_{L}$ gaugino mass and $m_{\ell_{i}}, i=1,2,3$ are the physical $e,\mu,\tau $ masses.
In our numerical results, however, the exact form of both $G^L_{\tilde t_{1} b\ell_{i}}$ and $G^R_{\tilde t_{1} b\ell_{i}}$ will be used.

The various parameters entering the vacuum expectation values (\ref{B1}),(\ref{B2}) and the partial widths (\ref{B4}) come in two classes, those--such as  $Y_{b}$, $Y_{t}$, $m_{\ell_{i}}$ and the gauge coupling $g_{2}$--that are physically measured quantities whose values we simply insert, and the rest, which form a large parameter space over which one must scan. Of this latter type, there are a number of constraints which relate them--such as demanding unification of the $g_{3}$, $g_{2}$, $g_{R}$ and $g_{BL}$ gauge couplings with related implications for the gaugino masses. Another set of constraints is directly related to the fact that the spontaneous breaking of $R$-parity occurs as a sneutrino VEV--thus linking the LSP decays to the neutrino mass matrix. In this paper, we will impose the condition that the LSP decays be ``prompt''--that is, well within the detection chamber at the LHC. It then follows that the dominant contribution to neutrino masses must be Majorana. 

The Majorana mass matrix can be computed in the minimal $B-L$ MSSM and is found to be
\begin{equation}
	\label{B7}
	{m_\nu}_{ij} = A {v_L}_i^* {v_L}_j^* + B \left({v_L}_i^* \epsilon_j + \epsilon_i {v_L}_j^* \right) + C \epsilon_i \epsilon_j \ ,
\end{equation}
where $A$, $B$ and $C$ are complicated flavor-independent functions of the above parameters.
As a first step, it is important to notice that the determinant of the neutrino mass matrix in (\ref{B7}) is zero. This is a consequence of the flavor structure and is independent of the $A, B$ and $C$ parameters. Closer observation reveals that only one eigenstate is massless. This constrains the neutrino masses to be either in the ``normal'' hierarchy (NH)
\begin{equation}
	m_1 = 0 < m_2 \sim 8.7 ~\text{meV} < m_3 \sim 50 ~\text{meV}
	\label{B8}
\end{equation}
or in the ``inverted'' hierarchy (IH)
\begin{equation}
	m_1 \sim  m_2 \sim 50 ~\text{meV} > m_3 = 0 \ .
\label{B9}
\end{equation}
In (\ref{B8}) and (\ref{B9}) we  have inserted $m_{1}=0$ and $m_{3}=0$ respectively into the squared mass differences measured in neutrino oscillation experiments and presented, for example, in \cite{Tortola:2012te, GonzalezGarcia:2012sz, Fogli:2012ua}. The constraints on the intitial parameters arise from diagonalizing (\ref{B7}) and inserting these values for the  neutrino masses, as well as the measured central values for the neutrino mixing angles--see, for example, \cite{Tortola:2012te, GonzalezGarcia:2012sz, Fogli:2012ua}. It is important to note that the central values for all of these mixing angles are determined with the exception of $\theta_{23}$. The data is consistent with this taking either one of two values--$\sin^{2}(\theta_{23})=0.587$ or  $\sin^{2}(\theta_{23})= 0.446$. In all cases, this class of constraints eliminates five of the six parameters $\epsilon_{i}, v_{L_{i}}$, $i=1,2,3$. We use the convention that the remaining unconstrained parameter is one of the $\epsilon_{i}$'s. Be this as it may, the precise constraining equations are different in each of the four cases: NH with $\sin^{2}(\theta_{23})=0.587$ or  $\sin^{2}(\theta_{23})= 0.446$ and IH with $\sin^{2}(\theta_{23})=0.587$ or  $\sin^{2}(\theta_{23})= 0.446$.

All of the above constraints reduce the number of independent parameters down to seven. Furthermore, demanding that the analysis should be ``generic'' without excessive fine-tuning of any parameters--as well as imposing lower bounds on some particle masses set by the LHC--limits the ranges of these parameters. The seven parameters, as well as their allowed ranges, are shown in Table 1.
\begin{table}[htdp]
\begin{center}
\begin{tabular}{|c|c|}
\hline
\hspace{1cm} Parameter \hspace{1cm} & \hspace{0.7cm} Range \hspace{0.7cm}
\\
\hline
$M_3$ (TeV) 	& 1.5 \, -- \, 10
\\
$M_{Z_R}$ (TeV)	&  2.5 \, -- \, 10
\\
$\tan \beta$ & 2 \, -- \, 55
\\
$\mu$ (GeV)& 150 \, -- \, 1000
\\
$m_{\tilde t_1}$ (GeV)& 400 \, -- \, 1000
\\
$\theta_{t}(\vphantom{t}^\circ)$ & 0 \, -- \, 90
\\
$\left|\epsilon_i\right|$ (GeV) & $10^{-4}$ \, -- \, $10^{0}$
\\
$\arg(\epsilon_i)(\vphantom{t}^\circ)$ & 0 \, -- \, 360
\\
\hline
\end{tabular}
\end{center}
\caption{The independent parameters and their ranges. The neutrino sector leaves only one unspecified $R$-parity violating parameter, which is chosen to be $\epsilon_i$ where the generational index, $i$, is also scanned to avoid any biases.} 
\label{scan}
\end{table}

We now proceed to give the results of a numerical analysis of the decays in (\ref{B3})--that is, of a stop LSP into a bottom quark and charged leptons. The branching ratio is defined as

\begin{equation}
\label{B10}
	{\rm Br}(\tilde t_1 \to b \ell^+_i) \equiv \frac{\Gamma(\tilde t_1 \to b \ell^+_i)}{\sum \limits_{i=1}^3 \Gamma(\tilde t_1 \to b \ell^+_i)} 
\end{equation}
and using the relation
\begin{equation}
	{\rm Br}(\tilde t_1 \to b \, e^+) + {\rm Br}(\tilde t_1 \to b \, \mu^+) + {\rm Br}(\tilde t_1 \to b \, \tau^+)=1 \ ,
\label{B11}
\end{equation}
one needs to present a plot of only two of the branching ratios--which we choose to be ${\rm Br}(\tilde t_1 \to b \, e^+)$ and $ {\rm Br}(\tilde t_1 \to b \, \tau^+)$. These quantities are numerically calculated using (\ref{B4}) by scanning over the parameters and ranges shown in Table 1. Since these ranges do not, by themselves, gaurantee that the stop remains the LSP, an additional check is implemented in the scan to throw out any points for which the stop cannot be the LSP.
In addition, the detailed constraint equations involving the
$\epsilon_{i}$, $v_{L_{i}}$ parameters are different in each of the four cases involving the NH versus the IH, as well as the two different central values for $\theta_{23}$. The results are shown in Figure 1.

\begin{figure}[h]
\centering
	\includegraphics[scale=0.4]{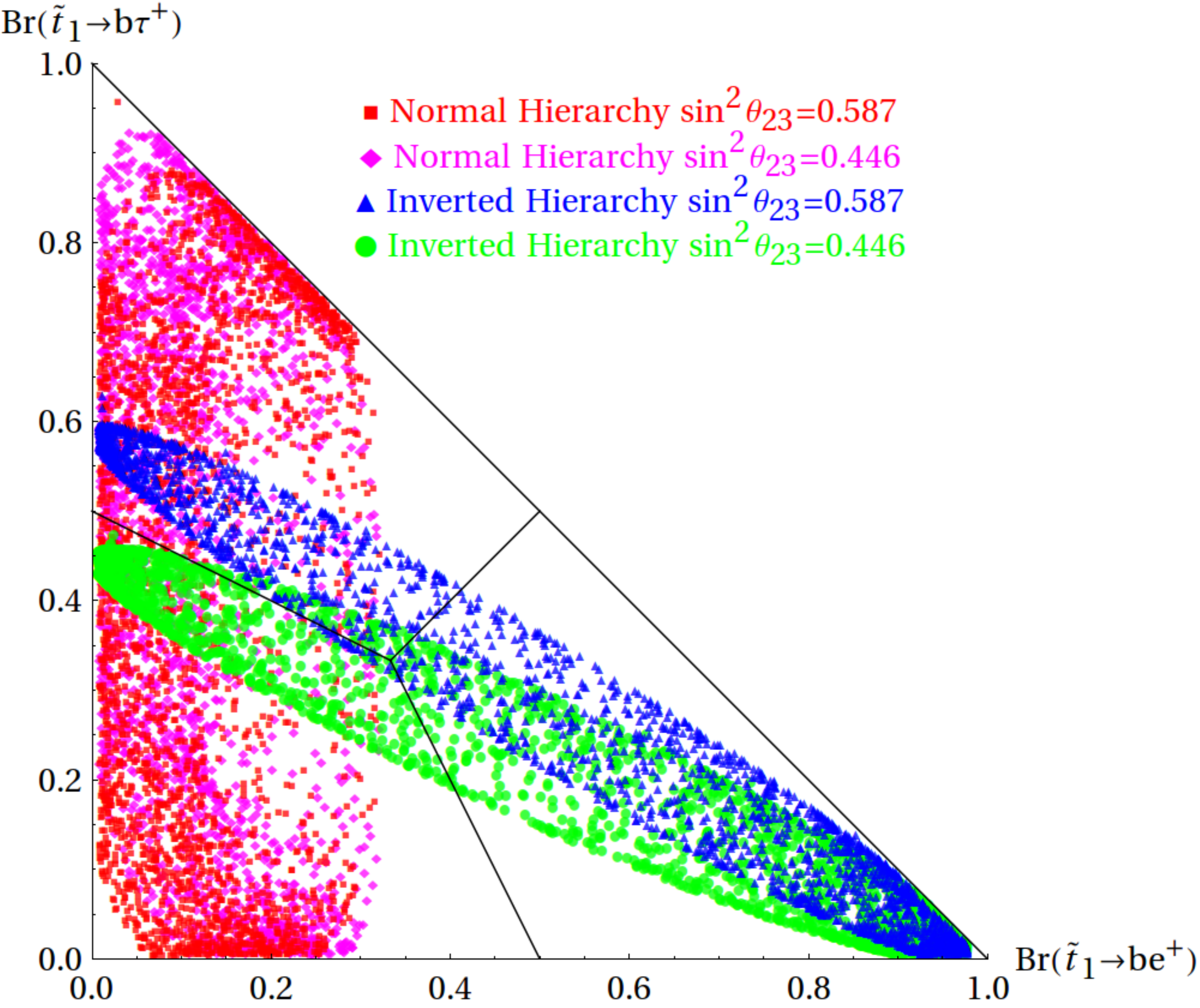}
	\caption{\footnotesize The results of the scan specified in Table~\ref{scan} using the central values for the measured neutrino parameters in the $\text{Br}(\tilde t_1 \to b \, \tau^+)$ - $\text{Br}(\tilde t_1 \to b \, e^+)$ plane. Due to the relationship between the branching ratios, the $(0,0)$ point on this plot corresponds to $\text{Br}(\tilde t_1 \to b \, \mu^+)=1$. The plot is divided into three quadrangles, each corresponding to an area where one of the branching ratios is larger than the other two. In the top left quadrangle, the bottom--tau branching ratio is the largest; in the bottom left quadrangle the bottom--muon branching ratio is the largest; and in the bottom right quadrangle the bottom--electron branching ratio is the largest. The two different possible values of $\theta_{23}$ are shown in blue and green in the IH (where the difference is most notable) and in red and magenta in the NH.}
	\label{fig:Brs.central}
\end{figure}

The conclusions to be drawn from Figure 1 are quite clear.

\begin{itemize}

\item If LHC data indicates bottom quark-charged lepton decays which intersect the populated region predicted by our numerical analysis, then a stop LSP of the minimal $B-L$ MSSM with the associated parameters is a distinct possibility. Were the LHC data to lie within the white regions of Figure 1, however, a stop LSP in this context is unlikely.

\item If the LHC data point lies in the top left quadrangle of Figure 1--where the bottom-tau branching ratio is the largest--then there are two possibilities. If the branching ratio to bottom-tau is highly dominant, then the neutrino masses are likely to be in the normal hierarchy and consistent with both values for $\sin^{2}(\theta_{23})$. On the other hand, if this branching ratio is only slightly dominant, then the data is compatible with both the normal and the inverted neutrino hierarchies. Were it to be shown by another experiment to be an inverted hierarchy, then this measurement would favor $\sin^{2}(\theta_{23})=0.587$ over $\sin^{2}(\theta_{23})=0.446$.

\item If the LHC data point lies in the bottom left quadrangle of Figure 1--where the bottom-muon branching ratio is the largest--then there are two possibilities. If the branching ratio to bottom-muon is highly dominant, then the neutrino masses are likely to be in the normal hierarchy and compatible with either value of $\sin^{2}(\theta_{23})$. On the other hand, if this branching ratio is only slightly dominant, then the data is compatible with both the normal and the inverted neutrino hierarchies. Were it to be shown by another experiment to be an inverted hierarchy, then this measurement would favor $\sin^{2}(\theta_{23})=0.446$ over $\sin^{2}(\theta_{23})=0.587$.

\item If the data point lies in the bottom right quadrangle--where the bottom-electron branching ratio dominates--then the neutrino masses are likely to be in an inverted hierarchy. If the data is in the upper part of the populated points, then this inverted hierarchy would be consistent with 
$\sin^{2}(\theta_{23})=0.587$. Data in the lower part of this region would indicate an inverted hierarchy with $\sin^{2}(\theta_{23})=0.446$.

\end{itemize}

\section*{Lower Bounds on the Mass of a Stop LSP}

Since a stop LSP in the minimal $B-L$ MSSM scenario decays as a leptoquark, one can set bounds on its mass using previous leptoquark searches at the LHC. Under the assumption in this paper that the stop LSP is an admixture, it decays predominantly into a bottom quark and a charged lepton. Stop LSP's are produced at the LHC in ${\tilde{t}}_{1}$-${\bar{\tilde{t}}_{1}}$ pairs, implying that the final state will consist of two jets and a pair of oppositely charged leptons. The current ATLAS and CMS analyses search for such final states assuming the oppositely charged leptons have the same flavor~\cite{Chatrchyan:2012st, Chatrchyan:2012sv, Chatrchyan:2012vza, ATLAS:2013oea,Aad:2011ch, ATLAS:2012aq, CMS:zva}\footnote{For interpretation of these results for stop decays in explicit trilinear $R$-parity violation see~\cite{Evans:2012bf}.}. This yields upper limits on the ${\tilde{t}}_{1}$-${\bar{\tilde{t}}_{1}}$ production cross section for each of the three possible flavors. The upper limit on the cross section is easily translated into a lower bound on the stop LSP mass, since the cross section depends only on the mass, and the center of mass energy, and falls off steeply as the mass increases.

\begin{figure}[h]
\center
\includegraphics[scale=0.4]{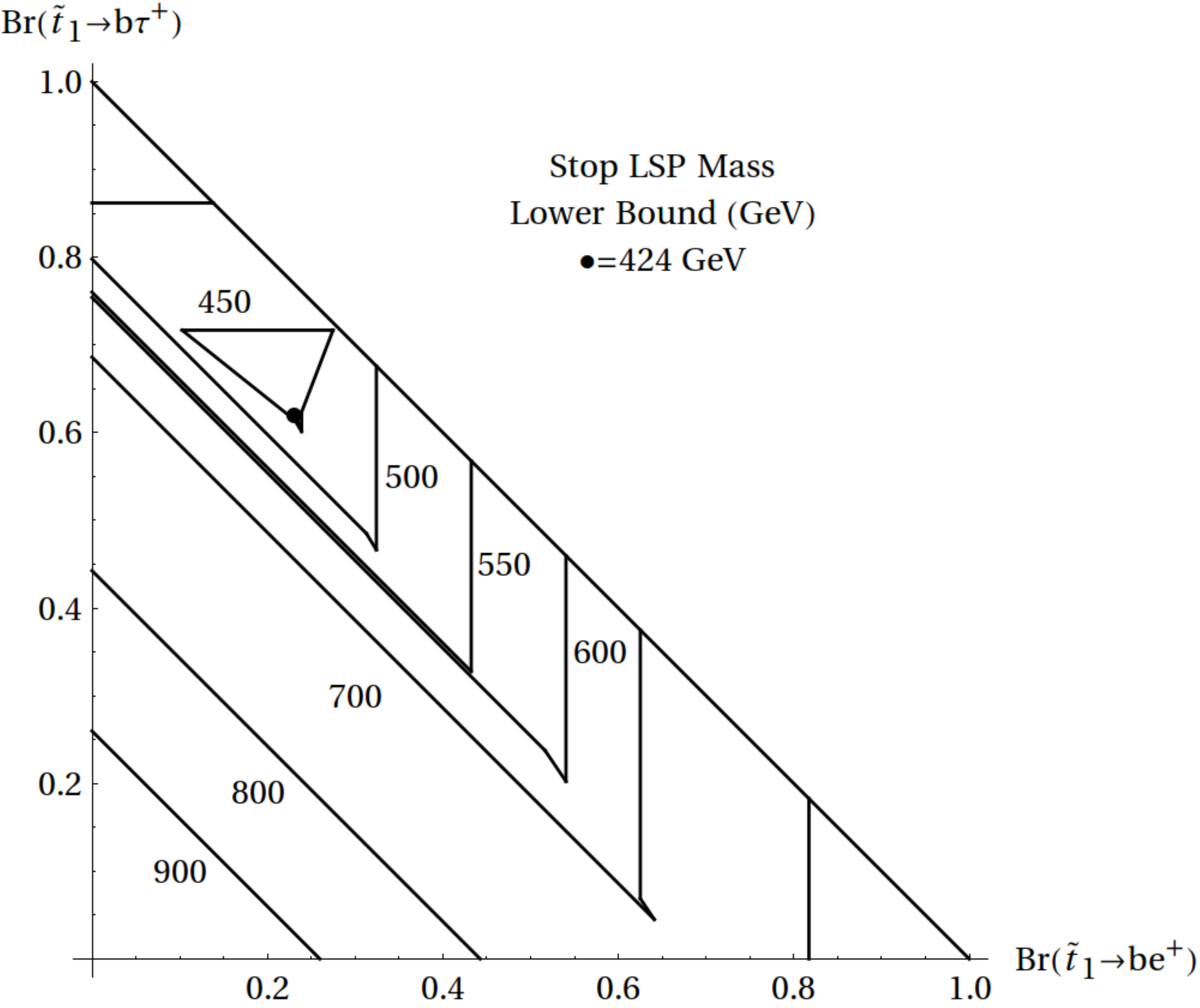}
\caption{\footnotesize Lines of constant stop lower bound in GeV in the $\text{Br}(\tilde t_1 \to b \, \tau^+)$ - $\text{Br}(\tilde t_1 \to b \, e^+)$ plane. The strongest bounds arise when the bottom--muon branching ratio is largest, while the weakest arise when the bottom--tau branching ratio is largest. The dot marks the absolute weakest lower bound at 424 GeV.}
\label{fig:stop.lower.bound}
\end{figure}

Although the ATLAS and CMS analyses assume branching ratios of unity to a given family, we can generalize their results to arbitrary branching ratios. This is accomplished by rescaling the cross section limit from each search by dividing it by the appropriate branching ratio squared. It is then compared to the calculated production cross section as a function of stop LSP mass, which yields the lower bound on the stop LSP mass from that search. For a given choice of branching ratios to $be^+$, $b\mu^+$, and $b\tau^+$, the search with the strongest expected stop mass lower bound is selected. Then the observed cross section limit from that search is rescaled in the same way and, finally, compared to the calculated production cross section as a function of stop LSP mass. This yields the lower bound on the stop LSP mass\footnote
{
	Experimental and background uncertainties place an approximate uncertainty on the stop mass lower bounds of~$\pm 50$ GeV in Figure~\ref{fig:stop.lower.bound}.
}. The production cross section, as calculated by the ATLAS, CMS and LPCC SUSY working group~\cite{Kramer:2012bx, Kramer2} at next-to-leading order in $\alpha_S$, including resummaiton at next-to-leading log, is used to place these lower bounds. Even though this cross section is calculated in the context of the $R$-parity conserving MSSM , it is valid here because the production cross section is dominated by $R$-parity conserving, color processes.

The exclusion results can, again, be plotted on a two-dimensional plot since the sum of all three branching ratios is unity. This is done in the form of lines of constant stop mass lower bound in Figure~\ref{fig:stop.lower.bound} in the $\text{Br}(\tilde t_1 \to b \, \tau^+)$ - $\text{Br}(\tilde t_1 \to b \, e^+)$ plane, the same plane as in Figure~\ref{fig:Brs.central}. The absolute lowest bound, 424~GeV, occurs at $\text{Br}(\tilde t_1\to be^+)=0.23$, $\text{Br}(\tilde t_1\to b \mu^+)=0.15$, $\text{Br}(\tilde t_1\to b \tau^+)=0.62$. It is marked by a dot. The bounds are stronger in the three corners of the plot where one of the branching ratios is unity. The strongest of these three bounds corresponds to decays purely to bottom--muon. This reflects the fact that this is the easiest of the three channels to detect and the search has been performed with the most integrated luminosity, 20 fb$^{-1}$, and center of mass energy, 8 TeV at CMS~\cite{CMS:zva}. The weakest of these bounds corresponds to decays purely to bottom--tau because this channel is the hardest to detect. The contours are each composed of several connected straight line segments. The straightness of the segments is due to the fact that the bound is always coming from a single channel (the one with the strongest expected bound) and, hence, only depends on one of the three significant branching ratios.

\newpage

\section*{Acknowledgments}
S.Spinner is indebted to P. Fileviez Perez for extensive discussion and a long term collaboration on related topics. S. Spinner would also like to thank the Max-Planck Institute for Nuclear Physics for hospitality during the early part of this work and T. Schwetz for useful discussion. B.A. Ovrut, A. Purves and S. Spinner are supported in part by the DOE under contract No. DE-AC02-76-ER-03071 and by the NSF under grant No. 1001296. The work of Z. Marshall is supported by the Office of High Energy Physics of the U.S. Department of Energy under contract DE-AC02-05CH11231.



\end{document}